\documentclass[twocolumn,showpacs,preprintnumbers,amsmath,amssymb]{revtex4}
\usepackage{graphicx}

\begin{document}
\preprint{}

\title{\bf Theoretical understanding  of the quasiparticle dispersion in bilayer high-$T_{c}$
 superconductors}

\author
{
 Jian-Xin Li$^{1,2,3}$, T. Zhou$^{1}$, and Z. D. Wang$^{1,2}$
}
\address{$^{1}$National Laboratory of Solid State Microstructures and
Department of Physics, Nanjing University, Nanjing 210093, China\\
$^{2}$Department of Physics, The University of Hong Kong, Pokfulam Road, Hong Kong, China\\
$^{3}$The Interdisciplinary Center of Theoretical Studies, Chinese
Academy of Sciences, Beijing, China. }
\date{\today}

\begin{abstract}

The renormalization of  quasiparticle (QP) dispersion in bilayer
high-$T_{c}$ cuprates is investigated theoretically by examining
respectively the interactions of the QP with spin fluctuations
(SF) and phonons. It is illustrated that both interactions are
able to give rise to a kink in the dispersion around the antinodes
(near $(\pi,0)$). However,  remarkable differences between the two
cases are found for the peak/dip/hump structure in the lineshape,
the QP weight, and  the interlayer coupling effect on the kink,
which are suggested to serve as a discriminance to single out the
dominant interaction in the superconducting state. A comparison to
recent photoemission experiments shows clearly that the coupling
to the spin resonance is dominant for the QP around antinodes in
bilayer systems.

\end{abstract}

\maketitle

 The elucidation of many-body interactions in high-$T_{c}$
superconductors (HTSC) is considered as an essential step toward
an insightful understanding of their superconductivity.
Angle-resolved photoemission spectroscopy (ARPES) has provided a
powerful way to probe the coupling of charge QPs to other QPs or
collective modes. Recent ARPES experiments unveil several
intriguing  features in the dispersion, the QP weight, and the
lineshape: (i) A kink in the dispersion was observed in both the
nodal and antinodal regions~\cite{lan,bori,kim,ding}. (ii) The QP
weight around the antinodal region decreases rapidly with the
reduction of dopings~\cite{Dam}, while changes a little around the
nodal direction ~\cite{johnson}. (iii) After disentangling the
bilayer splitting effect, an intrinsic peak/dip/hump (PDH)
structure was seen around the antinodal region both in the bonding
(BB) and antibonding (AB) band~\cite{bori,kim}. (iv) The kink
around the antinodal region seemly shows a different momentum,
temperature and doping dependence from that around the nodal
direction~\cite{bori}. (v) The antinodal kink is likely absent (or
very weak) in the single-layered Bi2201~\cite{ding}. These
features, especially (i), imply  that the QP is coupled to a
collective mode. So far, two collective modes of 41 meV spin
resonance ~\cite{fon,bourges} and $\sim$ 36 meV
phonon~\cite{cuk,dev} have been suggested, but which one is a key
factor responsible for the kink is  still much debatable
~\cite{lan,bori,kim,ding,nor,cuk,dev}. So, it raises an important
question as if the electronic interaction alone is responsible for
the intriguing QP dispersion, or to what extent the antinodal
(nodal) QP's properties are determined by the strong electronic
interaction.

 In this Letter, we not only answer the above important question but also
  present a coherent understanding on the above features by
 studying  in detail the respective effects of
the fermion-SF interaction and fermion-phonon interaction on the
QP dispersion based on the slave-boson theory of the bilayer
$t-t^{\prime}-J$ model. We find that though both couplings are
able to give rise to the kink structure near the antinodal region
in the QP dispersion, they differ remarkably in the following
aspects. (a) The lineshape arising from the spin resonance
coupling exhibits a clear PDH structure, while the phonon coupling
would lead to a reversed PDH structure, namely the peak is in a
larger binding energy than the hump. (b) The former coupling
causes a rapid drop of the QP weight near the antinodal region
with underdoping, but the latter  has only a very weak effect.
Moreover, the corresponding coupling constant for the fermion-SF
interaction is reasonable consistent with ARPES data, in contrast
to the much smaller value for the fermion-phonon interaction.
(c) The bilayer coupling plays a positive role in  the occurrence
of the kink in the case of the fermion-SF interaction, but has a
negative effect on the formation of the kink for the
fermion-phonon interaction. These results suggest that the SF
coupling be a dominant interaction involved in the
antinode-to-antinode scattering.

We will consider separately the interactions of fermions with the
SF and phonons.  Let us start with the bilayer $t-t^{\prime}-J$
model with the AF interaction included,
\begin{eqnarray}
H&=&-\sum_{<ij>,\alpha,\alpha',\sigma}t_{\alpha,\alpha'}
c^{(\alpha)\dagger}_{i\sigma}c^{(\alpha')}_{j\sigma}-
\sum_{<ij>',\alpha,\sigma}t^{'}c^{(\alpha)\dagger}_{i\sigma}c^{(\alpha)}_{j\sigma}
\nonumber \\
& &-h.c.+\sum_{<ij>\alpha,\alpha'}J_{\alpha,\alpha'}{\bf
S^{(\alpha)}_i}\cdot {\bf S^{(\alpha')}_j},
\end{eqnarray}
where $\alpha=1,2$ denotes the layer index, $t_{\alpha,\alpha'}=t,
J_{\alpha,\alpha'}=J$ if $\alpha=\alpha'$, otherwise, $
t_{\alpha,\alpha'}=t_{p}/2, J_{\alpha,\alpha'}=J_{\perp}/2$ and
$i=j$. Other symbols are standard. In the slave-boson
representation, the electron operators $c_{j\sigma}$ are written
as $c_{j\sigma}=b^{\dagger}_{j}f_{j\sigma}$, where fermions
$f_{i\sigma}$ carry spin and bosons $b_i$ represent the charge.
Using the mean field parameters
$\chi_{ij}=\sum_{\sigma}<f^{\dagger}_{i\sigma}
f_{j\sigma}>=\chi_0$, $\Delta_{ij}=<f_{i\uparrow}
f_{j\downarrow}-f_{i\downarrow} f_{j\uparrow}>=\pm\Delta_0$, and
setting $b\rightarrow \sqrt{\delta}$ with $\delta$ the doping
density (boson condense), we can decouple the Hamiltonian (1). Its
Fourier transformation is given by, $H_{m}=\sum_{{\bf
k}\sigma\alpha}\varepsilon_{\bf k}f^{(\alpha)\dagger}_{{\bf
k}\sigma}f^{(\alpha)}_{{\bf k}\sigma} -\sum_{{\bf
k}\alpha}\Delta_{\bf k}(f^{(\alpha)\dagger} _{{\bf
k}\uparrow}f^{(\alpha)\dagger}_{-{\bf k}\downarrow}+h.c.)
+\sum_{{\bf k}\sigma}[\delta t_{\perp{\bf
k}}e^{ik_zc}f^{(1)\dagger}_{{\bf k}\sigma}f^{(2)}_{{\bf
k}\sigma}+h.c.]+\varepsilon_0$, with $\varepsilon_{\bf
k}=-2(\delta t+J'\chi_0)(\cos k_x+\cos k_y)-4\delta t'\cos k_x
\cos k_y-\mu_f$, $\Delta_k=2J'\Delta_0(\cos k_x-\cos k_y)$,
$\varepsilon_0=4NJ'(\chi^{2}_0+\Delta^{2}_0)$, $t_{\perp{\bf
k}}=t_p(\cos k_x-\cos k_y)^{2}/4$~\cite{and} and $J'=3J/8$.
Diagonalizing the Hamiltonian, we get the AB and BB bands with the
dispersion $\epsilon^{(A,B)}=\varepsilon_{\bf k} \pm \delta
t_{\perp{\bf k}}$. Then, the bare normal (abnormal) Green's
functions of fermions $\mathcal{G}_s^{(A,B)}$
($\mathcal{G}_w^{(A,B)}$), and the bare spin susceptibility
$\chi^{\alpha,\alpha^{\prime}} (\alpha,\alpha^{\prime}=A,B)$ are
obtained. The physical spin susceptibility is given by
$\chi=\chi^{+}_{0}\cos^{2}(q_{z}c/2)+\chi^{-}_{0}\sin^{2}(q_{z}c/2)$,
with $\chi^{+}_{0}=\chi^{AA}+\chi^{BB}$ and
$\chi^{-}_{0}=\chi^{AB}+\chi^{BA}$.

The slave-boson mean-field theory underestimates the AF
correlation~\cite{lee,li}, so we need to go beyond it and include
the effect of SF through the random-phase approximation (RPA), in
which the renormalized spin susceptibility is,
\begin{equation}
\chi^{\pm}({\bf q},\omega)={\chi^{\pm}_0({\bf
q},\omega)}/[1+(\alpha J_{\bf q}\pm J_\perp)\chi^{\pm}_0({\bf
q},\omega)/2]
\end{equation}
However, in the ordinary RPA ($\alpha=1$), the AF correlation is
overestimated as indicated by a larger critical doping density
$\delta\approx 0.22$ for the AF instability than the experimental
data $\delta_{c}=0.2\sim 0.5$~\cite{kampf}. Thus, we use the
renormalized RPA in which the parameter $\alpha$ is determined by
setting the AF instability at the experimental value $\delta_{c}$.
The fermionic self-energy coming from the SF coupling   is given
by,
\begin{eqnarray}
\Sigma^{(A,B)}_{s,w}({\bf k})&=&\pm \frac{1}{4N\beta}\sum_{\bf
q}[(J_{\bf q}+J_\perp)^{2}\chi^{+}({\bf
q})\mathcal{G}^{(A,B)}_{s,w}({\bf
k}-{\bf q}) \nonumber \\
& & +(J_{\bf q}-J_\perp)^{2}\chi^{-}({\bf
q})\mathcal{G}^{(B,A)}_{s,w}({\bf k}-{\bf q})],
\end{eqnarray}
where, the $+(-)$ sign is for the normal (abnormal) self-energy
$\Sigma_{s(w)}$, and the symbol ${\bf q}$  represents an
abbreviation of ${\bf q},i\omega_{m}$. The renormalized Green's
function is $G^{(A,B)}({\bf k},i\omega)=[(G^{(A,B)}_{s}({\bf
k},i\omega))^{-1}+(\Delta_{{\bf
k}}+\Sigma^{(A,B)}_w)^{2}G^{(A,B)}_{s}({\bf k},-i\omega)]^{-1}$
with
$G_{s}^{(A,B)}=[i\omega-\epsilon^{(A,B)}-\Sigma^{(A,B)}_{s}]^{-1}$,
and the spectral function is obtained via
$A(k,\omega)=-(1/\pi)\mathrm{Im}[G(k,\omega+i\delta)]$. To
determine the QP weight $z$, we first write the Green's function
as $G({\bf
k},i\omega)=M/(i\omega+\Sigma_\alpha)+N/(i\omega-\Sigma_\beta)$
and get $z$ via
$z=M/[1+\partial\Sigma_\alpha/\partial\omega(\omega_0)]$, with
$\omega_0$ the pole of $G$. The parameters we choose are $t=2J$,
$t^{\prime}=-0.7t$, $J=120$ meV and $J_{\perp}=0.1J$~\cite{kampf}.
Except in Fig.4, the doping density is set at $\delta=0.2$.

\begin{figure}
\includegraphics[angle=0,width=8cm,totalheight=6.5cm]{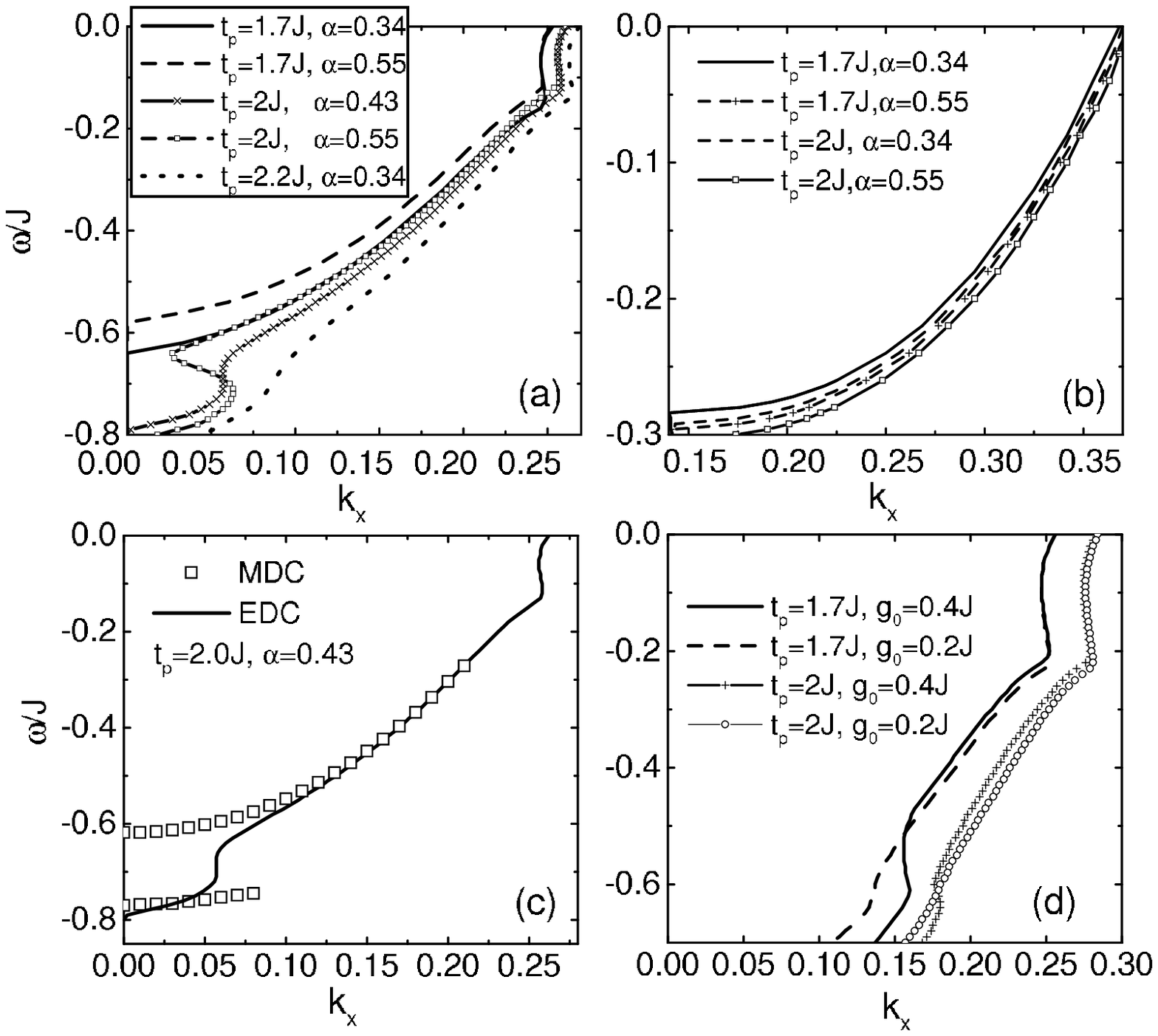}
\caption{\label{fig:epsart} The MDC dispersion of fermions due to
the interaction with spin fluctuations (a),(b) and (c), and
phonons (d). Figs.(a),(c) and (d) are the results at $(k_{x},\pi)$
(antinodal region), and Fig.(b) at $(k_{x},k_{x})$ (nodal
direction). The scattered points in Fig.(c) are derived from the
EDC~\cite{mdc}.}
\end{figure}
Fig.1 (a) and (b) display the calculated QP dispersion for the BB
band obtained from the momentum distribution curve
(MDC)~\cite{mdc} in the antinodal and nodal regions, respectively.
One can see from Fig.1(a) that the antinodal kink appears in some
range of parameters $t_{p}$ and $\alpha$, such as $t_{p}=2.0J$ and
$\alpha=0.43$. The RPA correction factor $\alpha=0.34,0.43$ and
0.55 correspond to the critical doping density of the AF
instability $\delta_{c}=0.02,0.05$ and $0.09$, while
$\delta_{c}=0.02\sim 0.05$ is within the experimental
range~\cite{kampf}. Meanwhile, the ARPES experiment indicated that
$t_{\perp,exp}$ (corresponding to $\delta t_{p}$, $\delta \sim
0.2$) is $44\pm 5$meV~\cite{feng}, i.e., $t_{p}=1.6\sim 2.0J$
here. Therefore, in the reasonable parameter range, the
interaction between fermions and SF reproduces well the observed
antinodal kink. In contrast, Fig.1(b) shows that no kink is
present in the nodal region. This can be understood from the
conservation of the momentum in the scattering process, namely the
nodal-to-nodal scattering involves a smaller transferred momentum
than ${\bf Q}=(\pi,\pi)$ where the AF spin fluctuation peaks.
Thus, we will focus mainly on the antinodal region in the
following discussion. In Fig.1(c) we replot the MDC dispersion
together with the EDC derived dispersion~\cite{mdc}. Near and
below the region where the kink appears in the MDC dispersion, the
EDC dispersion breaks into a two-part structure, a peak and a
hump. Therefore, the appearance of the antinodal kink has an
intimate relation to the PDH structure in the EDC plot, being in
agreement with experiments~\cite{nor2}.

In fact, the kink may also be expected if fermions are
predominantly coupled to other collective modes. A hotly discussed
mode responsible for the antinodal kink is the out-of-plane
out-of-phase O bucking $B_{1g}$ phonon~\cite{cuk,dev}. To take
into account this kind of mode, we may have the total Hamiltonian
by including the following interaction in the slave-boson mean
field Hamiltonian $H_{m}$,
\begin{equation}
H_{ep}=\frac{1}{\sqrt{N}}\sum_{{\bf k},{\bf q},\sigma}g({\bf
k},{\bf q})f^{\dagger}_{{\bf k},\sigma}f_{{\bf k}+{\bf
q},\sigma}(d^{\dagger}_{\bf q}+d_{-{\bf q}})
\end{equation}
where $d^{\dagger}$ and $d$ are the creation and annihilation
operators for phonons, $g({\bf k},{\bf q})=g_0[\Phi_x({\bf
k})\Phi_x({\bf k}-{\bf q})\cos (q_y/2)-\Phi_y({\bf k})\Phi_y({\bf
k}-{\bf q})\cos (q_x/2)]/\sqrt{\cos^{2}q_x/2+\cos^{2}q_y/2}$ and
the detailed form of $\Phi_x,\Phi_y$ can be found in
Ref.~\cite{cuk,dev}. We note that the vertex $g({\bf k},{\bf
q}=0)\sim \cos(k_{x})-\cos(k_{y})$ and vanishes for all ${\bf k}$
at ${\bf q}=(\pi,\pi)$~\cite{cuk,dev}. Thus, the fermions near
$(\pi,0)$ are strongly scattered by this interaction. The
corresponding  fermionic self-energy  is given by,
\begin{eqnarray}
\Sigma_{s,w}({\bf k})&=&\pm \frac{1}{\beta N}\sum_{\bf q}|g({\bf
k}-{\bf q},{\bf q})|^{2} \nonumber \\
& & D_{0}({\bf q})[\mathcal{G}^{(A)}_{s,w}({\bf k}-{\bf
q})+\mathcal{G}^{(B)}_{s,w}({\bf k}-{\bf q})].
\end{eqnarray}
where the Green's function of phonon is $D_{0}({\bf
q})=2\omega_{q}/[(i\omega)^{2}-(\omega_q)^{2}]$, and a
dispersionless optical phonon ($B_{1g}$) will be taken as $\hbar
\omega_q=36$meV~\cite{cuk,dev}. Unlike the fermion-SF interaction,
the coupling constant here is not available now. We have tried
various values and found that the well-established kink can be
obtained when $g_{0}\approx 0.2J\sim 0.4J$ if $t_{p}=1.7J$ as
shown in Fig.1(d). This result is quite similar to that for the SF
in Fig.1(a). So, a mere reproduction of the QP kink is not
sufficient to single out the main cause of the renormalization,
and we must resort to other features.

A meaningful difference between the effects of these two
interactions on the dispersion can be seen clearly from a
comparison of Fig.1(a) and (d), namely,  the interlayer coupling
$t_p$ has opposite impacts on the antinodal kink. It enhances the
kink feature for the fermion-SF interaction; as shown in Fig.1(a),
when $t_{p}$ decreases to  $1.7J$, no antinodal kink is observed
even for the strongest AF coupling $\alpha=0.55$ which is in fact
beyond the experimentally acceptable value. In contrast, the kink
feature is weakened by the interlayer coupling for the
fermion-phonon interaction; as seen in Fig.1(d),  the kink
disappears when $t_{p}$ increases to $2J$ from $1.7J$ with
$g_{0}=0.2J$. Recent ARPES experiments revealed that a more
pronounced kink is present in the multilayered BiSrCaCuO, in sharp
contrast to the case in the single-layered one~\cite{ding}, which
can be considered as an indication to favor the fermion-SF
interaction in the bilayer system. Because the spin susceptibility
in the bilayer system involves scatterings between layers, the
self-energy [Eq.(3)] has a feature that the fermions in the BB
(AB) band is scattered into AB (BB) band via the SF in the odd
channel $\chi^{-}$. The so-called spin resonance (a sharp peak in
Im$\chi$) appears around ${\bf Q}$ both in the odd $\chi^{-}$ and
even $\chi^{+}$ channels (see Fig.2(a)). However, it is more
prominent in the odd channel, which is in agreement with very
recent experiments~\cite{pailhes}, mainly due to the larger vertex
$\alpha J_{\bf Q}-J_\perp$ ($J_{\bf Q}=-2J$) [Eq.(2)], compared to
$\alpha J_{\bf Q}+J_\perp$ in the even channel. On the other hand,
the AB band (and its associated flat band near ${\bf Q}$) is much
close to the Fermi surface compared to the BB band, as shown in
Fig.2(b). As a result, the fermionic self-energy for BB band is
large, and consequently the renormalization is strong. We have
indeed observed that the AB band is much less affacted by this
coupling, so no kink is present in this band (not shown here). As
increasing $t_{p}$, the splitting between the AB and BB band
pushes the flat portion of the AB band to be more close to the
Fermi level [Fig.2(b) and (c)], so enhances the
scattering~\cite{note}. However, for the coupling to phonons, the
AB and BB bands contribute  to the self-energy in the same
way[Eq.(5)]. In this case, though the increase of $t_{p}$
increases the contribution from the AB band, the decrease from the
BB band overcompensates that increase, and thus the self-energy
decreases with $t_{p}$ on the whole.
\begin{figure}
\includegraphics[angle=0,width=7.7cm,totalheight=3.2cm]{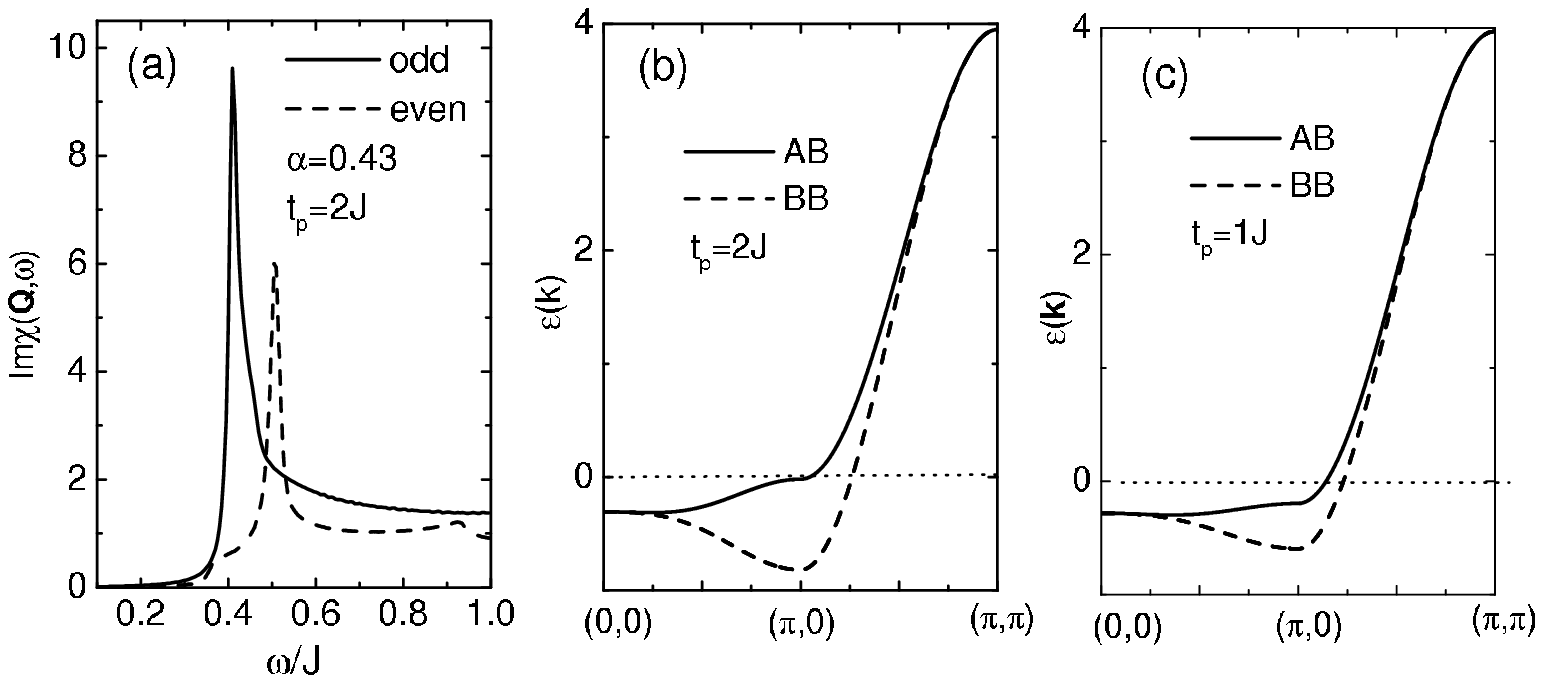}
\caption{\label{fig:epsart} (a) Im$\chi({\bf q},\omega)$ vs
$\omega$ at $(\pi,\pi)$. (b) and (c) show the bare dispersion of
the normal state quasiparticle for $t_{p}=2J$ and $1.0J$,
respectively.}
\end{figure}

\begin{figure}
\includegraphics[angle=0,width=7.6cm,totalheight=6.6cm]{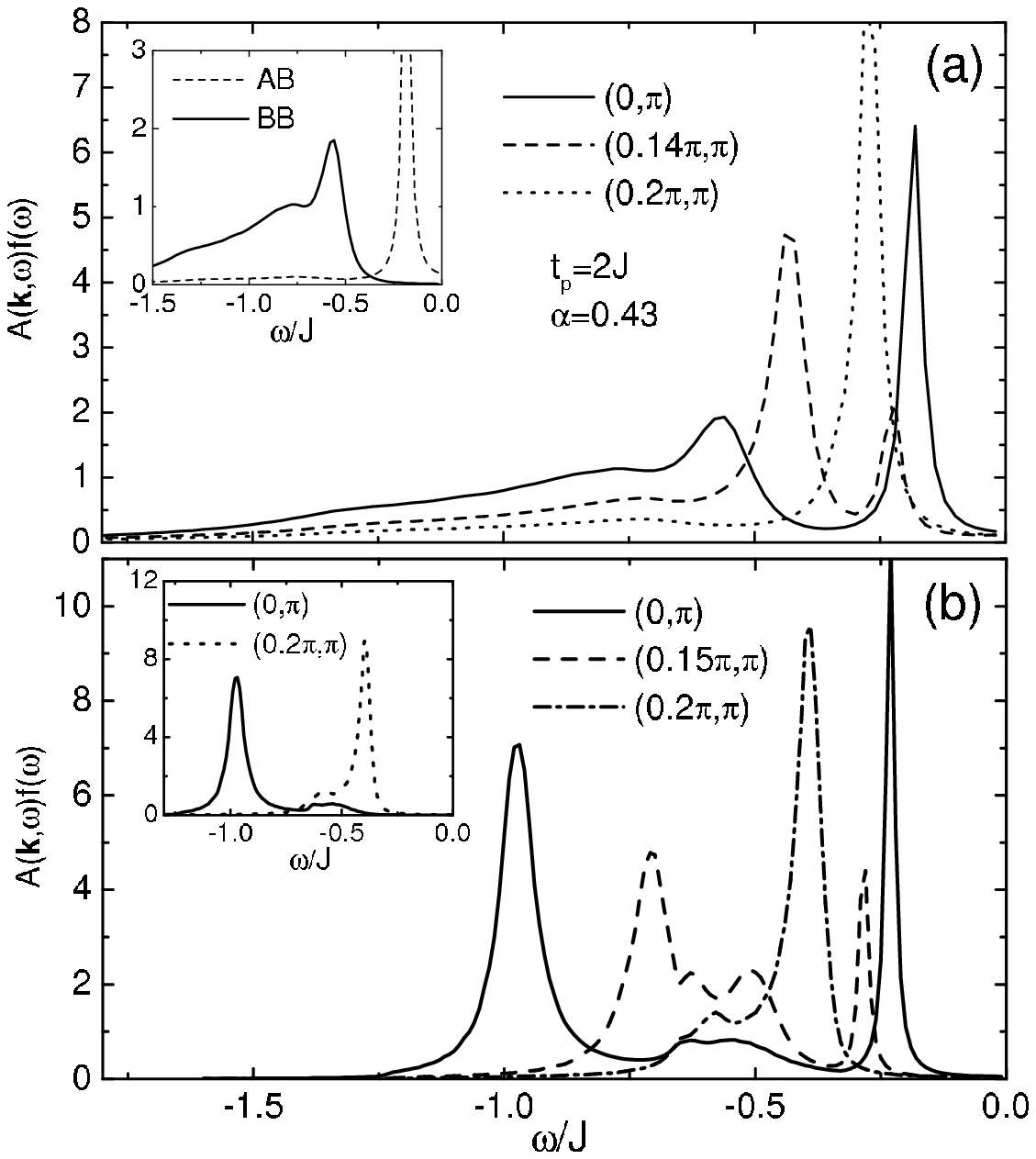}
\caption{\label{fig:epsart} The lineshape of fermions at different
$k$ points. Fig.(a) is obtained when fermions is coupled to spin
fluctuations. Fig.(b) shows the result arising from the coupling
to the $B_{1g}$ phonons. The inset in Fig.(a) shows the lineshape
for the bonding and antibonding bands at $(0,\pi)$, separately.
The inset in Fig.(b) is the lineshape for the bonding band at
${\bf k}=(0,\pi)$ and $(0.2\pi,\pi)$.}
\end{figure}
In Fig.3, we show the lineshape for different $k$ points from
$(0,\pi)$ to $(0,0.2\pi)$. For the fermion-SF interaction, both
the BB and AB spectra near $(0,\pi)$ consist of a low energy peak,
followed by a hump, and then a dip in between, though the
intensity of the AB hump is much weaker than its peak intensity
[inset of Fig.3(a)]; both develop their own PDH structure near
$(0,\pi)$. When moving from $(0,\pi)$ to $(0.2\pi,\pi)$, one will
see that the intensity of the BB peak increases, while the AB peak
decreases rapidly. Eventually, only the BB peak can be seen near
$(0.2\pi,\pi)$. These features are in good agreement with ARPES
experiments~\cite{feng}. However, the lineshape caused by the
phonon-coupling   displays a striking contrast to those shown in
Fig.3(a) and in experiments~\cite{feng} , namely, the peak is far
below the dip and the dip is below the hump [Fig.3(b)]. This is
because the renormalization to the QP peak in the BB band due to
the phonons is rather weak, so the peak is almost unchanged
comparing to the bare one. When moving from $(0,\pi)$ to
$(0.2\pi,\pi)$, the QP moves to be near the Fermi level, an
ordinary PDH structure is recovered [inset of Fig.3(b)] because
the hump arising from the phonon coupling does not change in the
process.

\begin{figure}
\includegraphics[angle=0,width=8cm,totalheight=6cm]{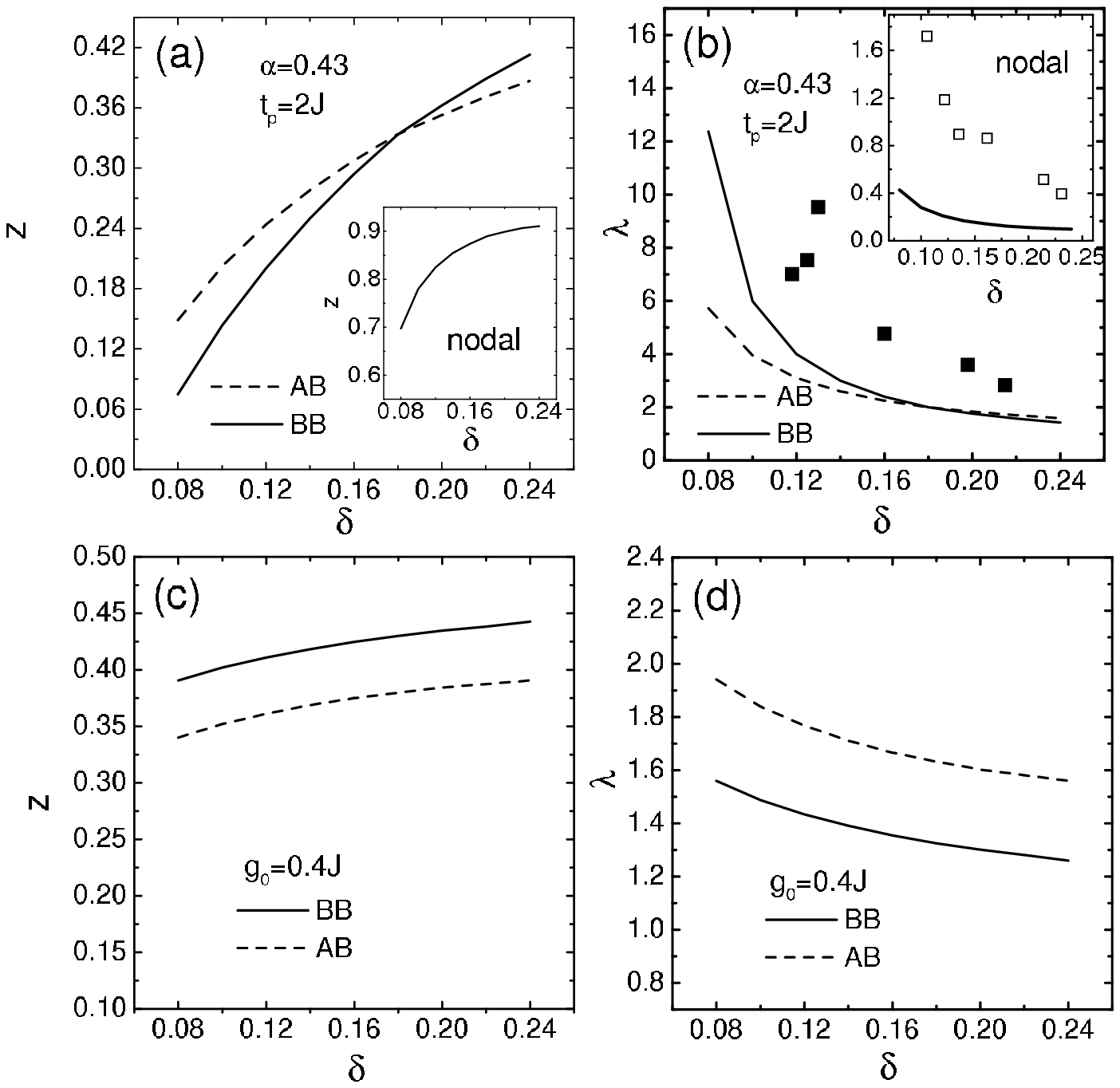}
\caption{\label{fig:epsart} The quasiparticle weight $z$ of
fermions and its coupling constant $\lambda$ at the antinodes
arising from the coupling to spin fluctuations (a) and (b), and to
phonons (c) and (d). The insets show those at the nodes. The
scattered points are experimental data from Ref.~\cite{kim} and
Ref.~\cite{johnson}.}
\end{figure}
Figure 4 presents the QP weight $z$ of fermions and the coupling
constant $\lambda$. Notice that the weight of the physical
electron is $\delta z$ due to the condensation of holons in the
slave-boson approach~\cite{lee}. For the fermion-SF interaction,
the weight decreases rapidly with underdoping, from nearly 0.42
and 0.39 at doping $\delta=0.24$ (the bare value is 0.5) to be
below 0.075 and 0.15 at $\delta=0.08$ for the BB and AB band,
respectively. On the other hand, the weight along the nodal
direction decreases very slowly, and it is still 0.7 even at
$\delta=0.08$. This exhibits the highly anisotropic interaction
between fermions and SF and is well consistent with what is
inferred from ARPES~\cite{Dam} and a recent argument based on the
analysis of experimental data~\cite{sheehy}. Moreover, the
coupling constant obtained using $z=1/(1+\lambda)$ shows a
reasonable fit to experimental data ~\cite{kim}[Fig.4(b)], while
that at the nodal direction is much smaller than the experimental
data~\cite{johnson} [inset of Fig.4(b)]. This consistence is
significant, because we use the well established parameters in the
$t-t^{\prime}-J$ model with only one adjustable parameter $\alpha$
being fixed by fitting to experiments. In contrast, the weight
decreases much slowly for the fermion-phonon interaction as shown
in Fig.4(c), and the coupling constant $\lambda$ is nearly $3\sim
5$ times smaller than experimental values.


Finally,
we wish to make two additional remarks. First, because the spin
resonance contributes little to the node-to-node scattering, a
kink structure in the nodal direction should be caused by the
other mode coupling, such as an in-plane Cu-O breathing phonon
~\cite{lan}. The different momentum, temperature and doping
dependence between the nodal and antinodal kink [feature (iv)]
supports this point of view. Second, since the interlayer coupling
plays opposite roles in the antinodal kink respectively for the
fermion-SF and fermion-phonon interactions, it is expected that,
even though a rather weak fermion-phonon coupling may be present
and lead to a weak antinodal kink-like behavior in the
single-layered cuprates, as possibly seen in the experiment for
Bi2201 ~\cite{ding}, the coupling is too weak to affect
significantly the lineshape which is mainly determined by the SF
and was shown to exhibit the peak/dip/hump structure in the
single-layered case~\cite{li}.

 We are grateful to F.C. Zhang for many useful discussions. The
work was supported  by the NSFC (10474032, 10021001, 10429401,
10334090), the RGC of Hong Kong (HKU 7050/03P), the URC fund of
HKU, and partly by RFDP (20030284008).


\begin{thebibliography}{99}

\bibitem{lan} A. Lanzara {\it et al.}, Nature {\bf 412}, 510
(2001).
\bibitem{bori} S. V. Borisenko {\it et al.}, Phys. Rev. Lett.
{\bf 90}, 207001 (2003); A. D. Gromko {\it et al.}, Phys. Rev. B
{\bf 68}, 174520 (2003).
\bibitem{kim} T.K. Kim {\it et al.}, Phys. Rev. Lett. {\bf 91},
167002 (2003).
\bibitem{ding} T. Sato {\it et al.}, Phys. Rev. Lett. {\bf 91},
157003 (2003).
\bibitem{Dam} A. Damascelli, Z. Hussain and Z. X. Shen, Rev. Mod.
Phys. {\bf 75}, 473 (2003).
\bibitem{johnson} P. D. Johnson {\it et al.}, Phys. Rev. Lett. {\bf 87},
177007 (2001).
\bibitem{fon} H. F. Fong {\it et al.},  Phys. Rev. Lett. {\bf 75}, 316 (1995).
\bibitem{bourges} P. Bourges {\it et al.}, Science {\bf 288}, 1234 (2000).
\bibitem{cuk} T. Cuk {\it et al.}, Phys. Rev. Lett. {\bf 93}, 117003 (2004).
\bibitem{dev} T. P. Devereaux, T. Cuk, Z. X. Shen, and N. Nagaosa, Phys. Rev. Lett. {\bf 93},
117004 (2004).
\bibitem{nor} M. Eschrig and M. R. Norman, Phys. Rev. Lett. {\bf
89}, 277005 (2002).
\bibitem{and} O. K. Anderson {\it et al.}, J. Phys. Chem. Solids {\bf 56}, 1573 (1995).
\bibitem{lee} P. A. Lee, Physica C {\bf 317-318}, 194
(1999); J. Brinckmann and P.A. Lee, Phys. Rev. Lett. {\bf 82},
2915 (1999).
\bibitem{li} J. X. Li, C. Y. Mou, and T. K. Lee, Phys.
Rev. B {\bf 62}, 640 (2000).
\bibitem{kampf} A. P. Kampf, Phys. Rep. {\bf 249}, 219 (1994).
\bibitem{mdc} Besides MDC, another intensity plot in ARPES is the energy
distribution curve (EDC). Due to the abnormal lineshape of
EDC~\cite{Dam}, MDC is more often used to analyze the dispersion.
\bibitem{feng} D. L. Feng {\it et al.}, Phys. Rev. Lett. {\bf 86},
5550 (2001).
\bibitem{nor2} M. R. Norman {\it et al.}, Phys. Rev. B {\bf 64},
184508 (2001).
\bibitem{pailhes} S. Pailhes {\it et al.}, cond-mat/0308394
(2003).
\bibitem{note} Notice that the interlayer AF coupling $J_{\perp}$
also enhances the coupling to spin resonance [see Eq.(3)].
\bibitem{sheehy} D. E. Sheehy, T. P. Davis, and M. Franz,
cond-mat/0312529 (2003).

\end{thebibliography}
\end{document}